\documentclass{article}

\usepackage{arxiv}

\usepackage[utf8]{inputenc} 
\usepackage[T1]{fontenc}    
\usepackage{hyperref}       
\usepackage{url}            
\usepackage{booktabs}       
\usepackage{amsfonts}       
\usepackage{nicefrac}       
\usepackage{microtype}      
\usepackage{lipsum}
\usepackage{graphicx}
\graphicspath{ {./images/} }

\title{General concept for autoignitive reaction wave covering from subsonic to supersonic regimes}

\author{
 Youhi Morii \\
  Institute of Fluid Science\\
  Tohoku University\\
  2-1-1 Katahira, Aoba, Sendai, Miyagi, 980-8577, Japan \\
  \texttt{morii@edyn.ifs.tohoku.ac.jp} \\
   \And
 Kaoru Maruta \\
  Institute of Fluid Science\\
  Tohoku University\\
  2-1-1 Katahira, Aoba, Sendai, Miyagi, 980-8577, Japan \\
}

\begin{document}
\maketitle
\begin{abstract}
We consider a one-dimensional (1D) autoignitive reaction wave in reactive flow system comprising unburned premixed gas entering from the inlet boundary and burned gas exiting from the outlet boundary.
In such a 1D system at given initial temperature, it is generally accepted that steady-state solutions can only exist if the inlet velocity matches either the velocity of deflagration wave, as determined by the burning rate eigenvalue in the subsonic regime or the velocity of detonation wave as dictated by the Chapman-Jouguet (CJ) condition in the supersonic regime.
In this study, we developed the general concept of "autoignitive reaction wave" and theoretically demonstrate that two distinct regimes that can maintain steady-state solutions both in subsonic and supersonic conditions.
Based on this theory, we selected inlet velocities that are predicted to yield either steady-state or flashback solutions, and conducted numerical simulations.
This novel approach revealed that steady-state solutions are possible not only at the velocity of the deflagration wave in the subsonic regime and the velocity of the detonation wave in the supersonic regime, but also across a broad range of inlet velocities.
Furthermore, we identify a highly stable "autoignitive reaction wave" that emerges when the inlet velocity surpasses the velocity of detonation wave, devoid of the typical shock wave commonly seen in detonation waves.
This "supersonic autoignitive reaction wave" lacks the instability-inducing detonation cell structure, suggesting the potential for the development of novel combustor concepts.
\end{abstract}

\section{Introduction}

Understanding the behavior of one-dimensional (1D) autoignitive reaction wave is critical for research and development using reactive flows.
In this study, we investigate a system characterized by unburned premixed gas entering from the inlet boundary and burned gas exiting from the outlet boundary.
Traditionally, it has been understood that a steady-state solution exists in the subsonic regime only when the inlet velocity matches the velocity of the deflagration wave, as determined by the burning rate eigenvalue \cite{Ferziger1993, Poinsot2005, Law, Williams}.
However, this conventional wisdom is predicated on the assumption that chemical reactions in the preheat zone are negligible.

Recent studies emphasize the significance of autoignition-assisted flames, which necessitate the inclusion of chemical reactions in the preheat zone \cite{Zhang2020, Ju2021, Gong2021}.
These flame behaviors are highly dependent on the ignition Damköhler number, defined as the ratio of the flow residence time to the ignition delay time.
In essence, the presence of autoignition-assisted flames suggests a plethora of steady-state solutions, contingent upon the computational domain size, which in turn affects the residence time of "preheat zone."
Building on these findings, we have successfully bridged the gap between 0D ignition phenomena and 1D deflagration waves theoretically \cite{Morii2022, Morii2023, Morii2023ICDERS}.
According to the theory, 0D ignition and 1D deflagration waves become equivalent after a spatio-temporal transformation, if a fuel Lewis number is unity \cite{Morii2022, Morii2023, Morii2023ICDERS}.
We introduce the term "autoignitive reaction wave" to describe a wave influenced by ignition yet behaving like a deflagration wave, as derived from the theory.
Contrary to the prevailing view that only a single steady-state solution exists for deflagration waves in subsonic 1D systems, our approach posits an infinite number of such solutions as "autoignitive reaction waves," asserting that ignition and flame are intrinsically linked.

We further extend the theory to scenarios involving supersonic inlet velocities.
In the supersonic regime, the conventional understanding is that a steady-state solution is possible only when the inlet velocity matches the detonation wave velocity, as dictated by the Chapman-Jouguet (CJ) conditions.
However, given that the "autoignitive reaction wave" originates from 0D ignition within "preheat zone," we argue that it should be independent of the inlet velocity.
Consequently, we propose that an infinite number of steady-state solutions exist for the "autoignitive reaction wave," even in supersonic conditions.

The primary objective of this study is twofold: to formally define the concept of "autoignitive reaction wave" and to delve into its theoretical foundations.
To achieve this, we examine the stability of the "autoignitive reaction wave" within a 1D system where the unburned premixed gas flows at velocities ranging from subsonic to supersonic.
In doing so, we uncover a previously unexplored stable wave structure within the system.

\section{Revisited of "Reaction Wave" Theory}

The intricate relationship between 0D ignition and 1D deflagration for a unity fuel Lewis number has been elaborated in our previous works \cite{Morii2022, Morii2023, Morii2023ICDERS}.  
Here, we focus exclusively on the theory pertinent to the "reaction wave."  

\subsection{Governing Equations}

We define normalized fuel mass fractions \( \tilde{Y}_\mathrm{f} \) and normalized temperature \( \tilde{T} \) as follows:

\begin{equation}
\tilde{Y}_\mathrm{f} = \frac{Y_{\mathrm{f, 1}} - Y_\mathrm{f}}{Y_\mathrm{f, 1} - Y_\mathrm{f, 0}},
\end{equation}

\begin{equation}
\tilde{T} = \frac{T - T_\mathrm{0}}{T_\mathrm{1} - T_\mathrm{0}},
\end{equation}

where \( Y_\mathrm{f} \) represents the mass fraction, \( T \) is the temperature, and subscripts \( 0 \) and \( 1 \) indicate the initial and final values, respectively.  
For a constant-pressure, constant-enthalpy case with a one-step chemical reaction model, the governing equations for 0D homogeneous ignition can be expressed as:

\begin{equation}
  \frac{\mathrm{d}\tilde{Y_\mathrm{f}}}{\mathrm{d}t}
  = -\frac{\dot{\omega}_\mathrm{f}W_\mathrm{f}}{\rho(Y_\mathrm{f,1} - Y_\mathrm{f,0})},
  \label{eq:0d-mass-fractions}
\end{equation}

\begin{equation}
  \frac{\mathrm{d}\tilde{T}}{\mathrm{d}t}
  = -\frac{\sum_{k=1}^{K} \dot{\omega}_kh_k W_k}{c_p \rho(T_\mathrm{1} - T_\mathrm{0})}.
  \label{eq:0d-temperature}
\end{equation}

Here, \( t \) is time, \( K \) is the total number of chemical species, \( \dot{\omega}_k \) is the chemical production rate of the \( k^\mathrm{th} \) species, \( W_k \) is the molecular weight of the \( k^\mathrm{th} \) species, \( \rho \) is the mass density, \( h_k \) is the enthalpy of the \( k^\mathrm{th} \) species, and \( c_p \) is the mean specific heat.  
Furthermore, the following relationship holds:

\begin{equation}
  \frac{\mathrm{d}\tilde{T}}{\mathrm{d}t} = -\frac{\mathrm{d}\tilde{Y}_\mathrm{f}}{\mathrm{d}t}.
  \label{eq:independentTime1}
\end{equation}

This equation implies that \( \tilde{T} \) and \( \tilde{Y}_\mathrm{f} \) have a Legendre-transformable relationship.  
Therefore, if either \( \tilde{T} \) or \( \tilde{Y}_\mathrm{f} \) is determined, all other state variables can also be determined.  
For the remainder of this section, we focus solely on equation (\ref{eq:0d-temperature}).  
Multi-step chemical reaction models can also be applied, provided appropriate progress variables can be identified and a bijective relationship between the chosen variable and \( \tilde{T} \) exists.  
For multi-step chemical reaction models, Tsunoda et al. discussed suitable progress variables \cite{Tsunoda2023}.  

The residence time \( \tau_\mathrm{res} \) is defined as the total time a fluid parcel spends within a control volume:

\begin{equation}
\tau_\mathrm{res} = \int_{x_0}^{x} u^{-1} \mathrm{d}x,
  \label{eq:ResidenceTime}
\end{equation}

where \( u \) represents the fluid parcel's velocity, and \( x \) is its position.  
The total differential of equation (\ref{eq:ResidenceTime}) is  \( \mathrm{d}t = u^{-1} \mathrm{d}x \).  
Thus, equation (\ref{eq:0d-temperature}) can be reformulated as:

\begin{equation}
  \frac{\mathrm{d}\tilde{T}}{\mathrm{d}t}
  =  u \frac{\mathrm{d}\tilde{T}}{\mathrm{d}x},
  \label{eq:0d-temperature-tilde-residence}
\end{equation}

This equation encapsulates the governing equations for both 0D ignition and 1D reaction wave when the fuel Lewis number is unity.  
Therefore, equation (\ref{eq:0d-temperature-tilde-residence}) is the governing equation for "general reaction wave" when the fuel Lewis number is unity.
Moreover, if chemical reactions can be neglected ahead of the shock wave, the relationship in equation (\ref{eq:0d-temperature-tilde-residence}) remains valid behind the shock wave.

\subsection{Steady-State Solutions for "Reaction Wave"}

We consider a one-dimensional (1D) system wherein unburned premixed gas enters from the left inlet boundary, and burned gas exits from the right outlet boundary.  
Traditionally, it has been believed that a steady-state solution exists only when the inlet velocity matches either the velocity of the deflagration wave, derived from the burning rate eigenvalue \cite{Ferziger1993, Poinsot2005, Law, Williams}, or the velocity of the detonation wave, derived from the Chapman-Jouguet (CJ) conditions.  
Such steady-state solutions can be computed using Cantera \cite{Cantera} or the Shock and Detonation Toolbox \cite{Shepherd2021}.  
Generally, only these two types of steady-state solutions are recognized for 1D reactive flow systems.  
However, this study reveals the existence of additional steady-state solutions when considering the "autoignitive reaction wave."  

To elucidate this, we employ Equation (\ref{eq:0d-temperature-tilde-residence}) to model a 1D system where unburned premixed gas flows from the left inlet boundary into an air-filled domain and freely exits from the right outlet boundary.  
Ignition occurs when \( \tau_\mathrm{res} \) equals the ignition delay time \( \tau_\mathrm{ig} \).  
The position \( x_\mathrm{ig} \) where ignition occurs can be determined as:

\begin{equation}
  x_\mathrm{ig} = \int_0^{\tau_\mathrm{ig}} u \mathrm{d}t \approx u_\mathrm{in} \tau_\mathrm{ig} = u_\mathrm{in} \tau_\mathrm{res},
  \label{eq:disp2}
\end{equation}

where \( x_\mathrm{ig} \) denotes the ignition position, and \( u_\mathrm{in} \) is the inlet velocity.  
In the 1D system, a deflagration or detonation wave forms around \( x_\mathrm{ig} \) post-ignition.  
If the inlet velocity matches the velocity of either wave, a steady-state solution exists.  
Conversely, it is generally assumed that if the inlet velocity is slower than the wave velocity, the wave will propagate towards the inlet (flashback), and if faster, it will be pushed towards the outlet, resulting in no steady-state solution.  
Nonetheless, our study posits that steady-state solutions should exist near \( x_\mathrm{ig} \) even when the inlet velocity surpasses that of either wave.  
We refer to these solutions, which stem from ignition but exhibit flame-like behavior, as the "reaction wave."

\begin{figure}[htb]
	\begin{center}
	\includegraphics[width=12cm,keepaspectratio]{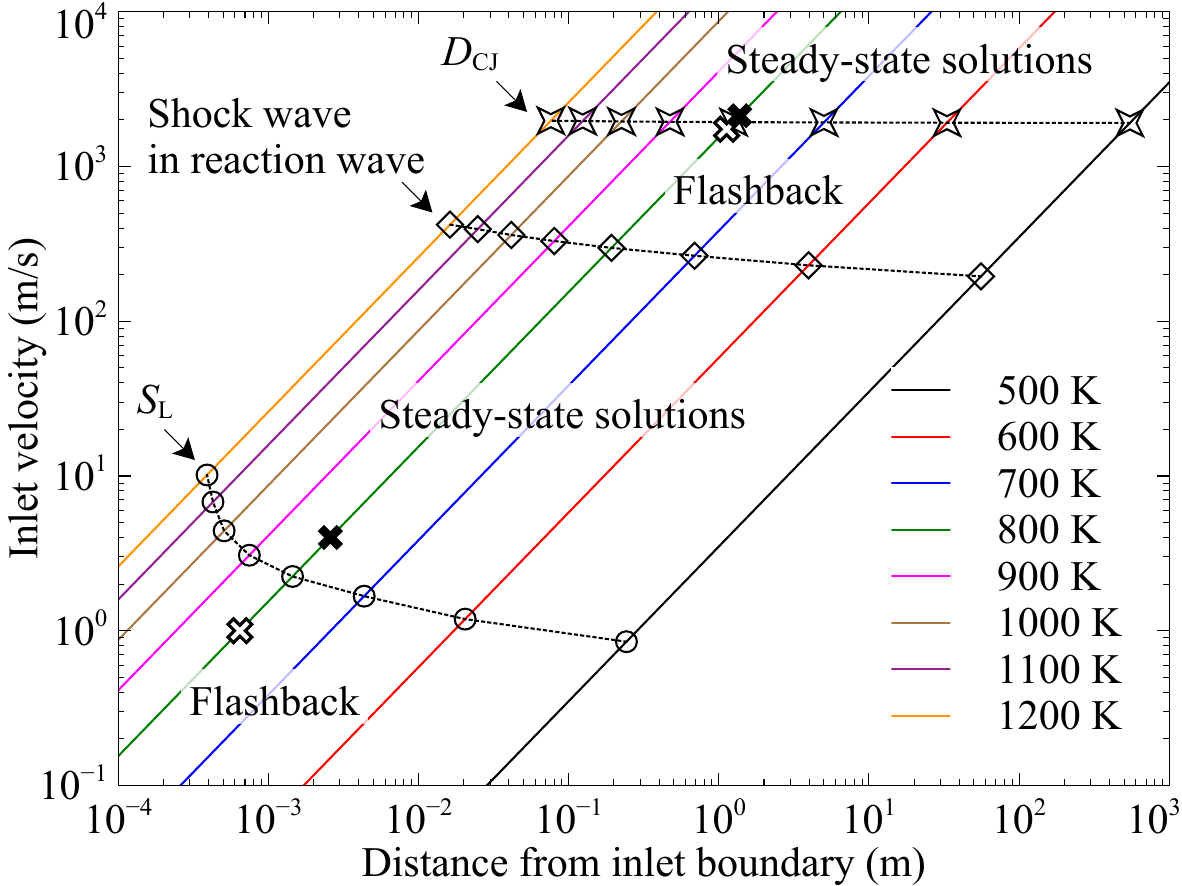}
	\end{center}
	\caption{
		The relationships between distance from inlet boundary and inlet velocity for stoichiometric methane-air mixtures with inlet temperatures varying from 500 to 1200 K at a pressure of 101325.0 Pa.
            The open circle-shaped markers mean that the inlet velocity is the same with the velocity of the deflagration wave.
            The open star-shaped markers mean that the inlet velocity is the same with the velocity of the detonation wave.
            The open diamond-shaped markers mean that the velocity can exceed the speed of sound within the "reaction wave" when the inlet velocity is subsonic.
		In addition, the open and close black crosses means the conditions where we computed in Sec. 4.
		}
	\label{fig-disp1}
	\vspace{2mm}
\end{figure}

Figure \ref{fig-disp1} illustrates the relationship between \( u_\mathrm{in} \) and \( x_\mathrm{ig} \) for stoichiometric methane-air mixtures with inlet temperatures varying from 500 to 1200 K at a pressure of 101325.0 Pa, using Equation (\ref{eq:disp2}).  
In the figure, open-circle markers indicate where the inlet velocity matches the velocity of deflagration wave, and open-star markers indicate a match with the velocity of detonation wave.  
This suggests that a "reaction wave" steady-state solution exists when the inlet velocity exceeds either of these marked velocities if autoignitive conditions are considered.  
However, since the detonation wave velocity is faster than that of the deflagration wave, this contradicts the conventional notion that no steady-state exists when the inlet velocity is slower than the velocity of detonation wave.  
This implies another transition point exists between the velocities of the two waves.

Given that the mass flow rate is constant, we have:

\begin{equation}
  \rho_\mathrm{in} u_\mathrm{in} \approx \rho u.
  \label{eq:disp3}
\end{equation}

This relationship can be rearranged into a correlation between temperature and velocity using the ideal gas equation, resulting in \( u \propto T \).  
However, the sound speed \( c \) is related to the temperature as \( c \propto \sqrt{T} \).  
Thus, in subsonic conditions, the velocity can surpass the sound speed within the "reaction wave," necessitating the formation of a shock wave.  
Beyond this transition point, no steady-state solutions exist as the detonation wave propagates towards the inlet boundary, causing a flashback.  
This transition is marked by open-diamond shapes in Figure \ref{fig-disp1}.  

In summary, this study predicts the existence of steady-state solutions not just at the two points where the inlet velocity matches the velocities of the deflagration and detonation waves, but also in two distinct regions.  
These regions are defined by inlet velocities that either exceed the velocity of deflagration wave without generating a shock wave within the "reaction wave," or surpass the velocity of detonation wave.
The steady-state solution for a given initial temperature depend on the ignition Damköhler number determined by both the distance from the inlet boundary and the inlet velocity. 
In addition, the steady-state solution could be related to mild combustion \cite{Cavaliere2004,Noman2016}.
We will proceed to validate these hypotheses through numerical simulations.

\section{Numerical Setups}

The objective of this study is to investigate the existence of steady-state solutions under conditions where the inlet velocity differs from the velocities of both deflagration and detonation waves.  
To achieve this, we developed an in-house compressible reactive flow solver named "myRNS," which leverages the high-performance programming language "Taichi Lang," primarily designed for computer graphics applications \cite{Taichi}.  
The solver utilizes the one-dimensional Navier-Stokes equations as governing equations, while neglecting the Dufour effect, the Soret effect, and pressure diffusion.

For transport coefficients, we optimized the Prandtl number \(Pr\) and the Lewis number \(Le_s\) of the \(s^{th}\) species using results from the Cantera software \cite{Cantera} prior to calculations.  
The viscosity coefficient \(\mu\) is determined by optimizing the coefficient \(C\) in Sutherland's formula \(\mu = \frac{T_0+C}{T+C}\left( \frac{T}{T_0}\right) ^{\frac{3}{2}}\), again using Cantera's results.  
Here, \( \mu_0 \) is the viscosity at \( T_0 = 273.15 \) K, as calculated using Cantera.  
The heat capacity at constant pressure \( C_p \) is evaluated in line with Cantera, while thermal conductivity \( \lambda \) and species diffusion coefficient \( D_s \) are calculated as \( \lambda = \frac{\mu C_p}{Pr} \) and \( D_s = \frac{\lambda}{\rho C_p Le_s} \), respectively.

\subsection{Numerical Methods}

Numerical fluxes are computed using the Kurganov and Tadmor central scheme \cite{Kurganov2000}, and higher-order spatial accuracy is achieved through the 5th-order WENO--Z+M scheme \cite{Luo2021m}.  
The viscous, heat conductivity, and diffusion terms are approximated using 4th-order central differences.  
Time integration is performed using a 3rd-order total variation diminishing (TVD) Runge--Kutta scheme \cite{Gottlieb1998}, while the chemical reaction time integration employs an explicit Euler method.  
The chemical reaction model is a one-step model, with constants based on Westbrook and Dryer's paper \cite{Westbrook1981}.

\subsection{Numerical Conditions}

Before initiating calculations, the computational domain was filled with air at a pressure of 101325.0 Pa having the same velocity as the inlet velocity. 
To enable comparison with our theoretical predictions, a stoichiometric methane-air mixture was used, as its fuel Lewis number is close to unity.  
The inlet boundary condition specifies both temperature and mass flow rate, while the outlet pressure is maintained at 101325.0 Pa.  
When the inlet velocity matches the velocity of either the deflagration or detonation wave, the existence of a steady-state solution is intuitively expected.  
However, elevated inlet temperatures can influence both types of waves.  
The velocity of deflagration wave \( S_{\mathrm{L}} \) and the velocity detonation wave \( D_{\mathrm{CJ}} \) at an inlet temperature of 800.0 K are 2.24 and 1936.02 m/s, respectively, with an ignition delay time of 0.6475 ms.  
Four inlet velocities were selected for study: 0.5 and 2.0 times the velocity of deflagration wave, and 0.9 and 1.1 times the velocity of detonation wave.  
Consequently, the chosen inlet velocities are 1.12, 4.48, 1742.42, and 2129.62 m/s.  
The corresponding estimated \( x_{\mathrm{ig}} \) values, calculated using \( u_{\mathrm{in}} \tau_{\mathrm{ig}} \), are \( 0.73 \times 10^{-3} \), \( 2.90 \times 10^{-3} \), 1.13, and 1.38 m, respectively.  
These values are marked by open diamond-shaped markers in Fig. \ref{fig-disp1}.  
Here, open cross-shaped markers indicate predicted flashback results, while closed cross-shaped markers indicate steady-state solutions as predicted by our theory.  
The computational domain length \( \ell \) was chosen to be sufficiently large to capture the reaction waves including deflagration and detonation, approximately twice \( x_{\mathrm{ig}} \).  
The mesh size was set at 10 \( \mu \)m; tests with a 20.0 \( \mu \)m mesh confirmed the adequacy of this resolution.  
It should be noted that similar results were obtained under different inlet temperature conditions.

\begin{table}[htpb]
	\caption{Numerical conditions for each case and predicted results from our theory.
 The inlet temperature is 800.0 K, and the ambient pressure is 101325.0 Pa for all cases. The mesh size is $10 \mu \mathrm{m}$.}
	\begin{center}
		\begin{tabular}{ccccc}
			\hline
			Cases & \( u_{\mathrm{in}} \) (m/s) & \( \ell \) (m) & \( x_{\mathrm{ig}} \) (m) & Predicted results \\ \hline 
			\( 0.5 \times S_{\mathrm{L}} \) & 1.12 & 0.0020 & \( 0.73 \times 10^{-3} \) & Flashback \\
			\( 2.0 \times S_{\mathrm{L}} \) & 4.48 & 0.0080 & \( 2.90 \times 10^{-3} \) & Steady-state \\
			\( 0.9 \times D_{\mathrm{CJ}} \) & 1742.42 & 2.5 & 1.13 & Flashback \\
			\( 1.1 \times D_{\mathrm{CJ}} \) & 2129.62 & 2.5 & 1.38 & Steady-state \\ \hline
		\end{tabular}
	\end{center}
\end{table}

\begin{figure}[htpb]
	\begin{center}
		\includegraphics[width=6.0cm,keepaspectratio]{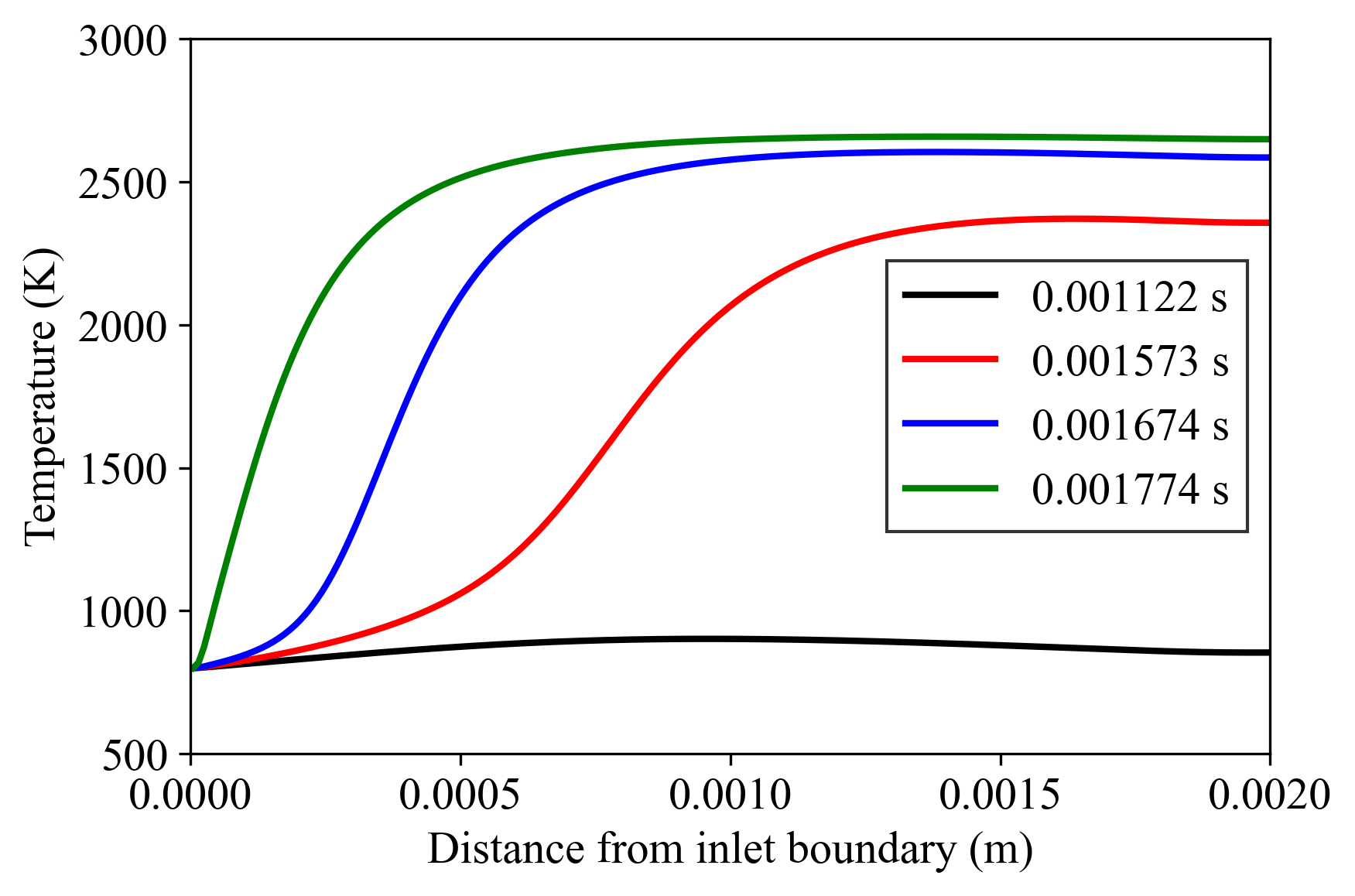} 
		\includegraphics[width=6.0cm,keepaspectratio]{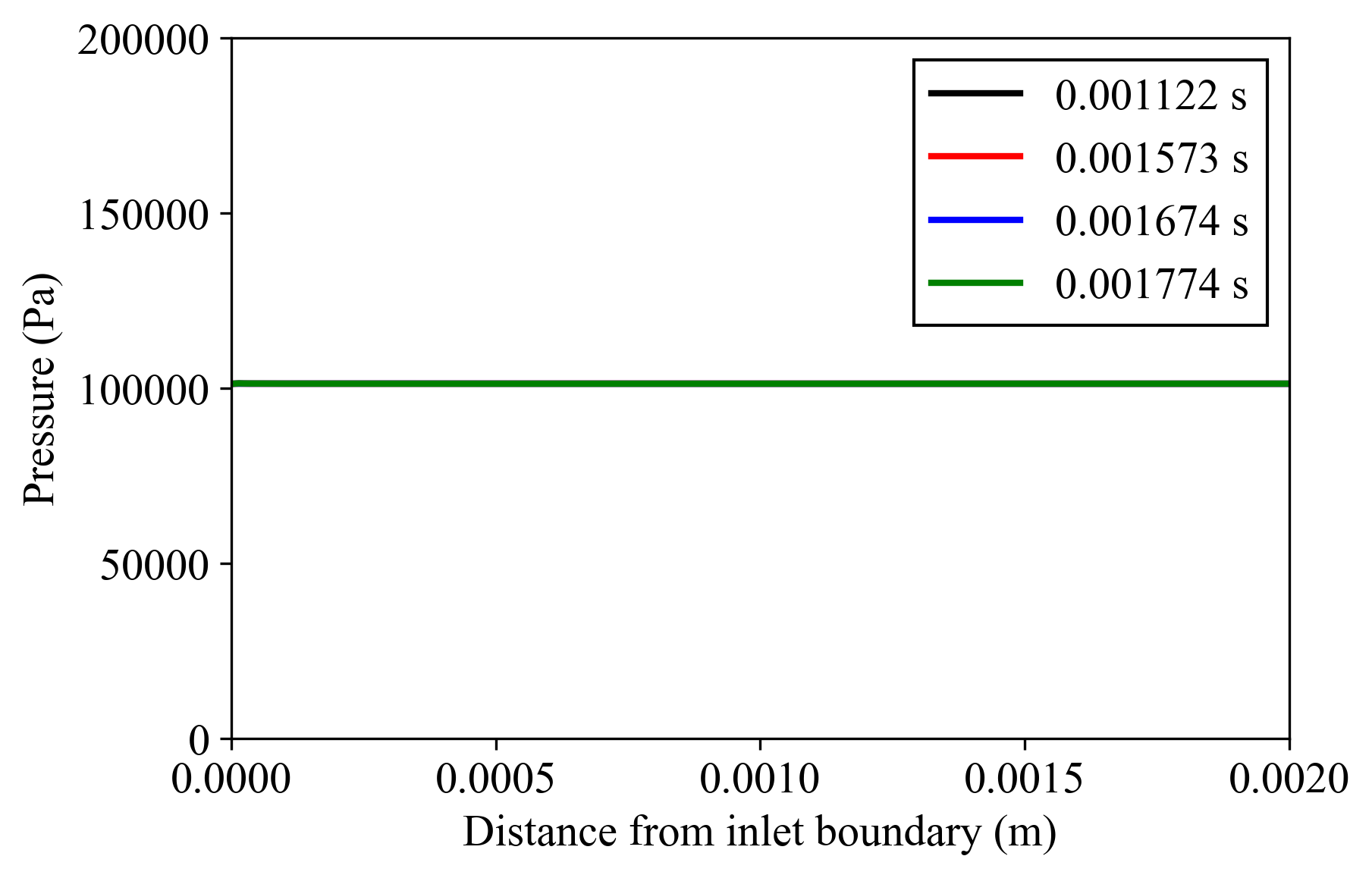} \\
		(a) $0.5 \times S_\mathrm{L}$ case, Flashback\\
		\includegraphics[width=6.0cm,keepaspectratio]{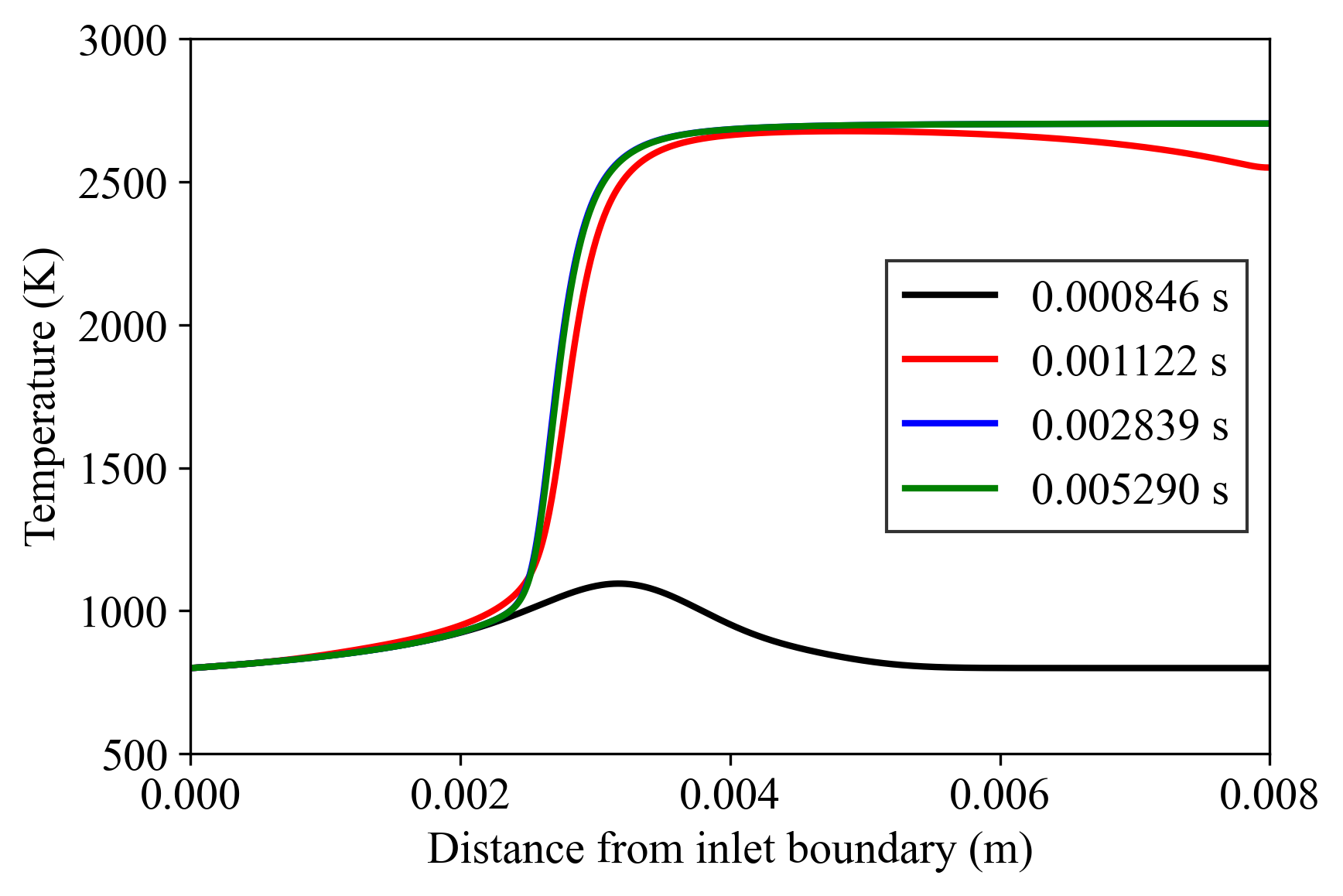} 
		\includegraphics[width=6.0cm,keepaspectratio]{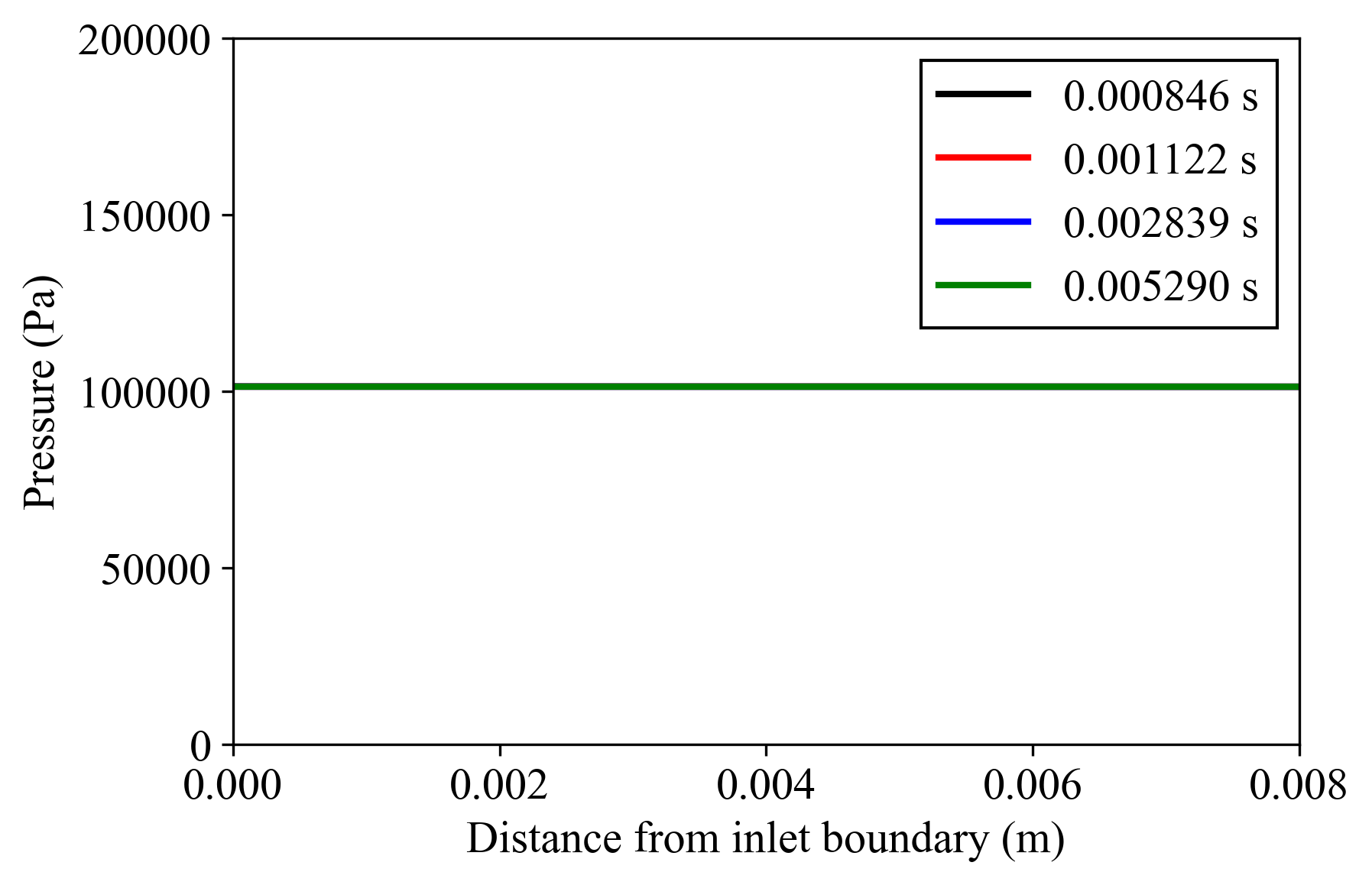} \\
		(b) $2.0 \times S_\mathrm{L}$ case, Steady-state\\
		\includegraphics[width=6.0cm,keepaspectratio]{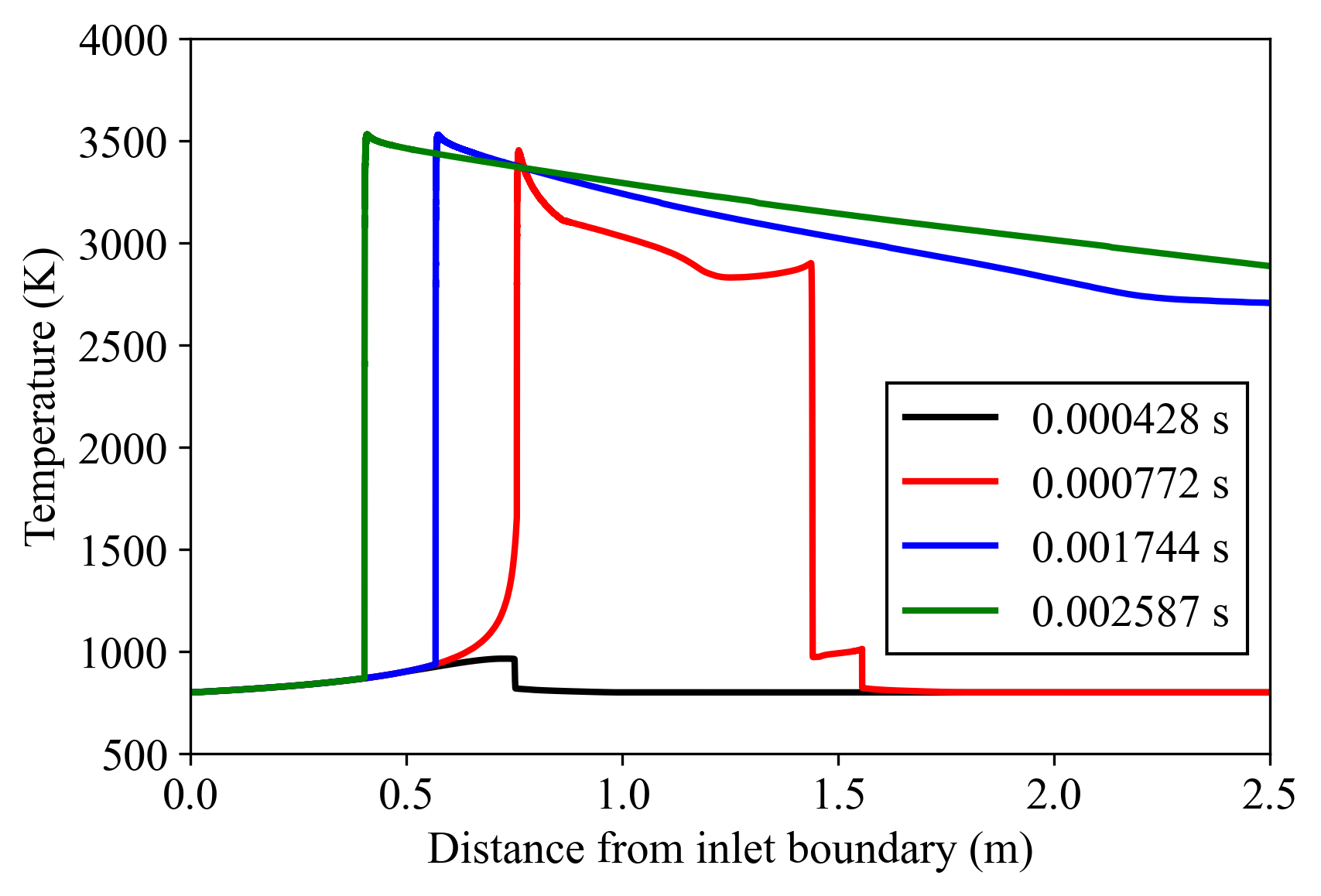} 
		\includegraphics[width=6.0cm,keepaspectratio]{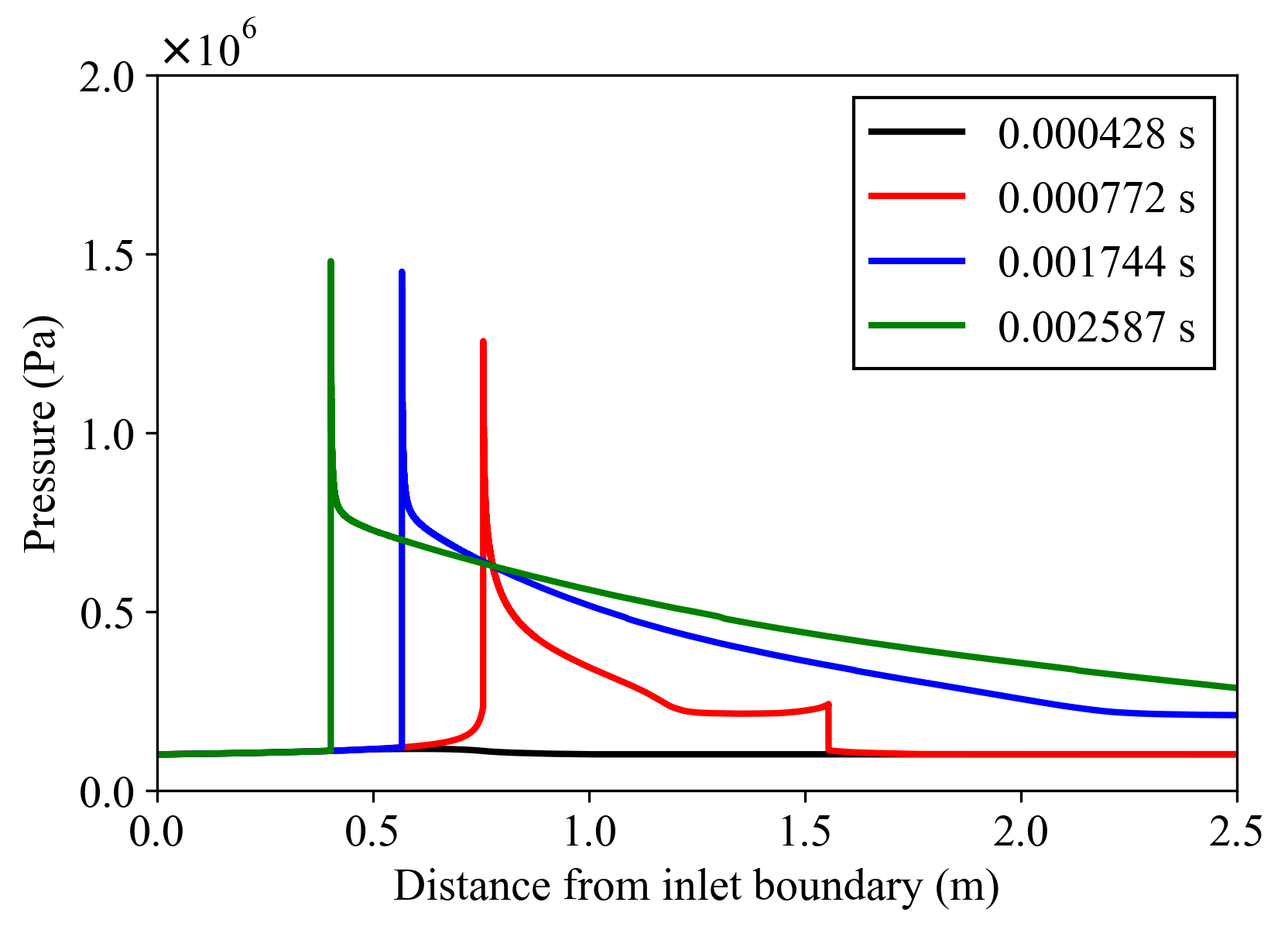} \\
		(c) $0.9 \times D_\mathrm{CJ}$ case, Flashback\\
		\includegraphics[width=6.0cm,keepaspectratio]{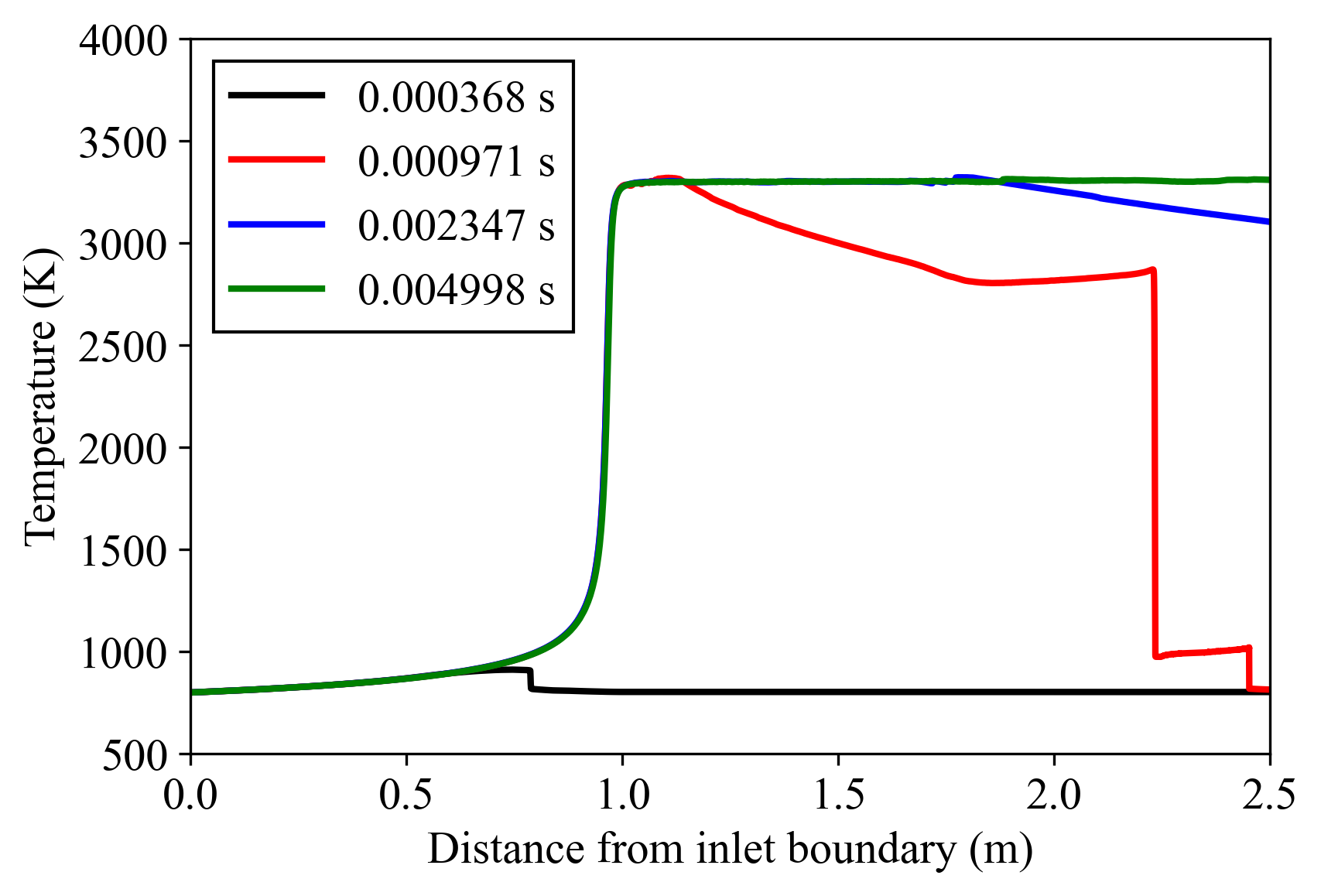} 
		\includegraphics[width=6.0cm,keepaspectratio]{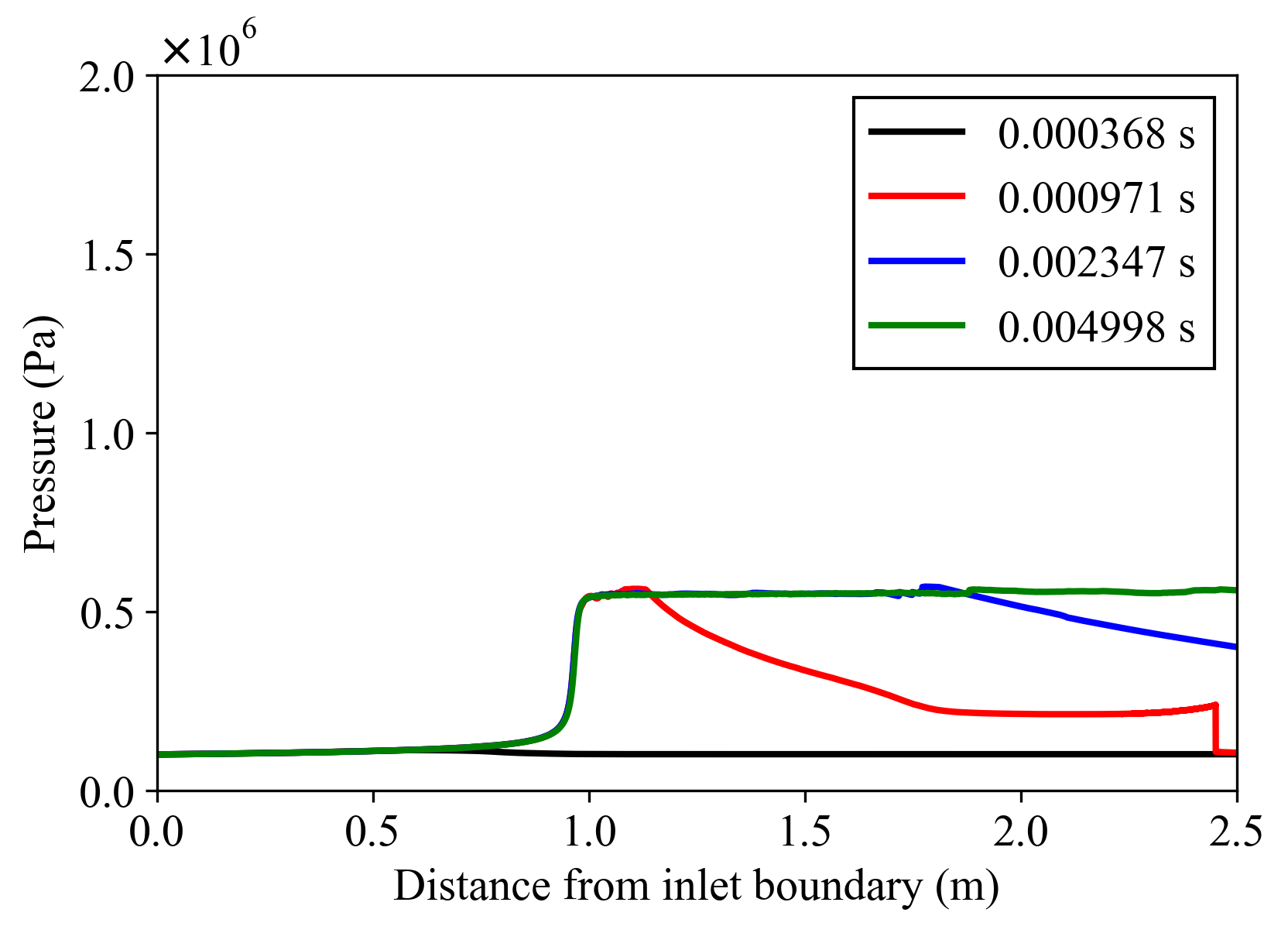} \\
		(d) $1.1 \times D_\mathrm{CJ}$ case, Steady-state
	\end{center}
	\caption{Instantaneous temperature profiles with four inlet velocities, (a) $0.5 \times S_\mathrm{L}$, (b) $2.0 \times S_\mathrm{L}$, (c) $0.9 \times D_\mathrm{CJ}$, and (d) $1.1 \times D_\mathrm{CJ}$ cases, are shown.Note that time passed the order of black, red, blue, and green in each figure.}
	\label{fig-disp2}
	\vspace{2mm}
\end{figure}
\section{Results and Discussion}

The outcomes of the simulations conducted under the conditions specified in Table 1 are illustrated in Figure \ref{fig-disp2}.  
As outlined in Section II, flashback phenomena were observed when the inlet velocity was smaller than \( S_\mathrm{L} \) (Fig. 2(a)) or \( D_\mathrm{CJ} \) (Fig. 2(c)).  
Conversely, the "reaction wave" remained stationary under conditions where the inlet velocity exceeded \( S_\mathrm{L} \) (Fig. 2(b)) or \( D_\mathrm{CJ} \) (Fig. 2(d)).  
Interestingly, no shock wave was detected when the inlet velocity was greater than \( D_\mathrm{CJ} \), and the profile resembled that at an inlet velocity close to \( S_\mathrm{L} \), albeit at higher equilibrium temperatures and pressures.  
Additionally, the absence of a shock wave structure and lower peak values for both temperature and pressure were noted when the inlet velocity was higher than \( D_\mathrm{CJ} \).  
In Table 1, the \( x_\mathrm{ig} \) values for \( 2.0 \times S_\mathrm{L} \) and \( 1.1 \times D_\mathrm{CJ} \) are \( 2.90 \times 10^{-3} \) and 1.38 m, respectively.  
These positions closely align with the locations of the reaction waves in Figures 2(b) and 2(d), albeit with minor discrepancies.  
For more precise results, \( x_\mathrm{ig} \) should be calculated using \( \int_0^{\tau_\mathrm{ig}} u \mathrm{d}t \) rather than approximated as \( x_\mathrm{ig} \approx u_\mathrm{in} \tau_\mathrm{ig} \).  
Nevertheless, even this approximation provides a reasonably accurate prediction of the stabilized position.

\section{Concluding Remarks}

The study reveals that steady-state solutions exist not merely at the two points where the inlet velocity matches the velocities of the deflagration or detonation waves, but also in a broader region if autoignitive conditions are considered.  
Notably, when the inlet velocity surpasses the detonation wave velocity, a stable reaction wave devoid of shock waves is observed.  
The peak temperature and pressure values within this "autoignitive reaction wave" are lower than those under detonation conditions.

\section*{Acknowledgement}
This work was partially supported by JSPS KAKENHI Grant Number 19KK0097.

\bibliographystyle{unsrt}  
\bibliography{library}
%
%
%
%
%
%

\end{document}